\documentclass[vecphys]{svmult}
\usepackage{graphicx}        
\usepackage{multicol}        
\usepackage[bottom]{footmisc}
\usepackage{amsmath,amsfonts}

\begin{document}

\DeclareGraphicsExtensions {.jpg,.pdf,.png}
\title*{Improvements to the Prototype Micro-Brittle Linear Elasticity Model of Peridynamics}
\titlerunning{Improvements to the PMB Model of Peridynamics}

\author{Georg C. Ganzenm\"uller\and
Stefan Hiermaier\and
Michael May}
\institute{Fraunhofer Ernst-Mach Institute for High-Speed Dynamics, Freiburg im Breisgau, Germany
\texttt{georg.ganzenmueller@emi.fraunhofer.de}}
\maketitle

\begin{abstract}~This paper assesses the accuracy and convergence of the linear-elastic, bond-based
Peridynamic model with brittle failure, known as the prototype micro-brittle (PMB) model.  We
investigate the discrete equations of this model, suitable for numerical implementation.  It is
shown that the widely used discretization approach incurs rather large errors.  Motivated by this
observation, a correction is proposed, which significantly increases the accuracy by cancelling
errors associated with the discretization. As an additional result, we derive equations to treat the
interactions between differently sized particles, i.e., a non-homogeneous discretization spacing.
This presents an important step forward for the applicability of the PMB model to complex
geometries, where it is desired to model interesting parts with a fine resolution (small particle
spacings) and other parts with a coarse resolution in order to gain numerical efficiency. Validation
of the corrected Peridynamic model is performed by comparing longitudinal sound wave propagation
velocities with exact theoretical results. We find that the corrected approach correctly reproduces
the sound wave velocity, while the original approach severely overestimates this quantity.
Additionally, we present simulations for a crack growth problem which can be analytically solved
within the framework of Linear Elastic Fracture Mechanics Theory. We find that the corrected
Peridynamics model is capable of quantitatively reproducing crack initiation and propagation.
\end{abstract}


\section{Introduction}
Peridynamics (PD), originally devised in 1999 by S. A. Silling \cite{GCGanzenmueller::Silling:2000a} is is a relatively new
approach to solve problems in solid mechanics. In contrast to the most popular numerical methods for
solving continuum mechanics problems, namely the Finite Element Method or the Finite Volume Method,
PD does not require a topologically connected mesh of elements.  Additionally, PD incorporates the
description of damage and material failure from the outset.  Within the context of mesh-free
methods, Peridynamics can be classified as a Total-Lagrangian collocation method with nodal
integration. PD features two classes of interaction models, so called bond-based materials and
state-based materials. In the bond-based case, interactions exist as spring-like forces between
pairs of particles. The interactions only depend on the relative displacement (and potentially its
history) of the interacting particle pair and are thus independent of other particles. This is in
contrast to the state-based model where pair-wise interactions also depend on the cumulative
displacement state of all other particles within the neighborhoods of the two particles which form
the pair.

\begin{figure}[!ht]
\begin{center}
\includegraphics[width=4in]{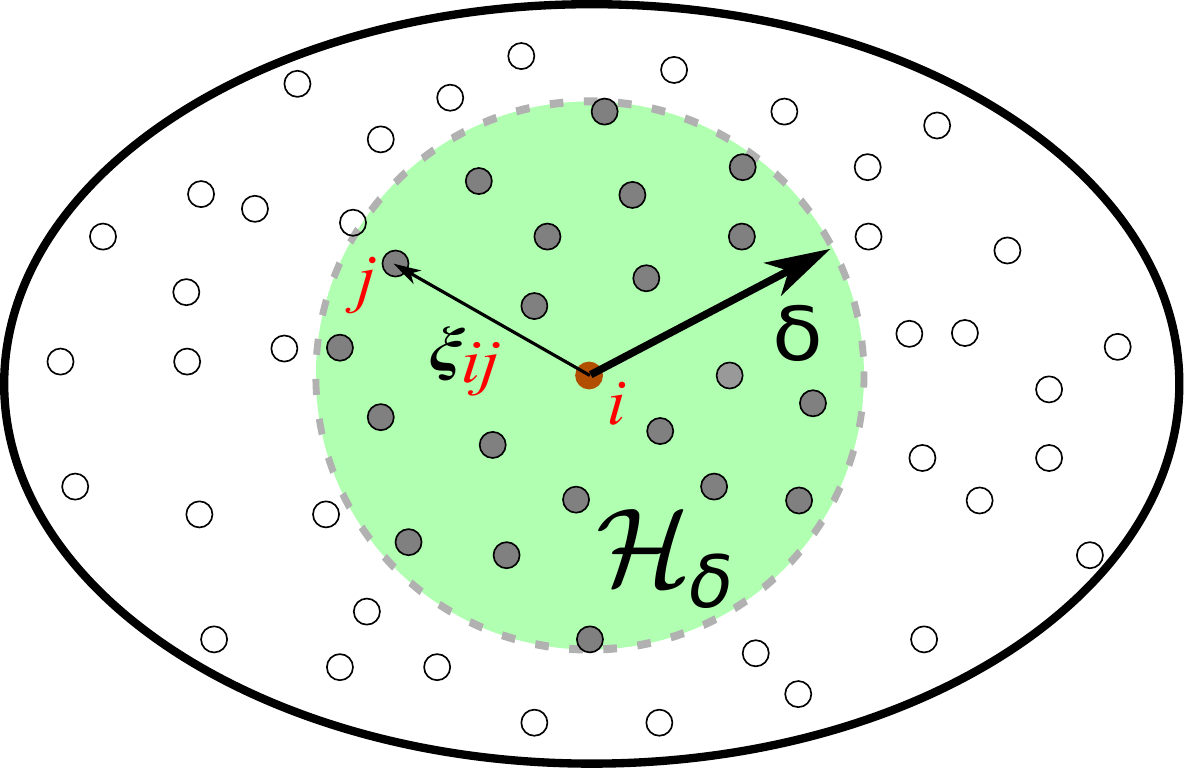}
\end{center}
\caption{
Peridynamics is a method for solving problems in solid mechanics. A body is discretized with a set
of integration nodes, which form the reference configuration.  Within this reference configuration,
each source node interacts with other nodes that are located within a finite horizon $\mathcal{H}_{\delta}$,
centered on the source node. The interactions are termed bonds. Peridynamics is a non-local method, because not only
nearest, or, adjacent, neighbors are considered.  The figure above depicts a single source node $i$
with a horizon given by the radial cutoff $\delta$. Bonds exist between node $i$ and all other nodes
$j$ which are inside $\mathcal{H}_{\delta}$. Upon deformation of the bonds, forces are projected along the
reference bond vectors $\vec{\xi}_{ij}$ such that solid material behavior is obtained.}
\label{GCGanzenmueller::fig:peridynamics_scheme}
\end{figure}

The scope of this paper is to assess the accuracy and convergence of the linear-elastic, bond-based
PD model with brittle failure, known as the prototype micro-brittle (PMB) model in the literature.
We investigate the discrete equations of this model, suitable for numerical implementation.
It is shown that the widely used discretization approach incurs rather large errors.
Motivated by this observation, a new discretization scheme is proposed, which significantly increases the
numerical accuracy. As an additional result, we derive equations to treat the interactions between
differently sized particles, i.e., a non-homogeneous discretization spacing. This presents an 
important step forward for the applicability of the PMB model to complex geometries, where
it is desired to model interesting parts with a fine resolution (small particle spacings) and other
parts with a coarse resolution in order to gain numerical efficiency.

We begin by introducing the basic terminology of bond-based PD.
In order to be consistent with the major part of the existing PD literature, we use the following
symbols: a coordinate in the reference configuration is denoted with $\vec{X}$, deformed (current)
coordinates are denoted by $\vec{x}$, such that the displacement is given by $\vec{u} = \vec{X} -
\vec{x}$. Bold mathematical symbols like the preceding ones denote vectors, while the same
mathematical symbol in non-bold font refers to its Euclidean norm, e.g. $x = |\vec{x}|$.

The governing equation for a PD continuum is given by
\begin{equation}
	W(\vec{X}, t) =  \frac{1}{2} \int \limits_{\mathcal{H}_{\delta}} 
        \omega(\vec{X}'-\vec{X})\,
        w \left[ \vec{u}(\vec{X}',t) - \vec{u}(\vec{X},t),	\vec{X}' - \vec{X} \right]\,
        \mathrm{d} V_{\vec{X}'},
	\label{GCGanzenmueller::eqn:peridynamic_energy_continuous}
\end{equation}
where $W(\vec{X}, t)$ is the energy density at a point located at $\vec{X}$ in the reference
configuration, and displaced at time $t$ by an amount $\vec{u}(\vec{X},t)$.\\  \mbox{$w \left [
\vec{u}(\vec{X}',t) - \vec{u}(\vec{X},t) \right ]$} is the \textit{micropotential}, which describes the
strain energy due to the relative displacement of a pair of points located at $\vec{X}$ and $\vec{X}'$. The
assumption that the strain energy density depends only on pairs of interacting volume elements leads
to the restriction of a fixed Poisson ratio of $1/3$ in 2D (1/4 in 3D).  The function
$\omega(\vec{X}'-\vec{X})$ is a weight function which modulates the pair interaction strength
depending on spatial separation, and $V_{\vec{X}'}$ is the volume associated with a point.

Referring to Fig.~\ref{GCGanzenmueller::fig:peridynamics_scheme}, the integration domain $\mathcal{H}_{\delta}$ is
the full disc (full sphere in 3D) around $\vec{X}$ described by the radial cutoff $\delta$, and is
termed the \textit{horizon}.  Within the PD picture, the strain energy is conceptually stored in
\textit{bonds} that are defined between all pairs of points $\left(\vec{X}, \vec{X}'\right)$ located within
$\mathcal{H}_{\delta}$. Thus, a bond vector in the reference configuration is given by $\vec{\xi} =
\vec{X}' - \vec{X}$, and the relative bond displacement due to some deformation at time $t$ is
$\vec{\eta}(t) = \vec{u}(\vec{X}',t) - \vec{u}(\vec{X},t)$.  The bond distance vector in the current
configuration is therefore written as $\vec{r}(t) = \vec{\eta}(t) + \vec{\xi}$.

With this notation, and dropping the explicit dependence on time, equation~(\ref{GCGanzenmueller::eqn:peridynamic_energy_continuous})
is written in a more compact form as
\begin{equation}
	W(\vec{X}) = \frac{1}{2}\int \limits_{\mathcal{H}_{\delta}}
	\omega(\vec{\xi})
        w(\vec{r}, \vec{\xi}) \mathrm{d} V_{\vec{X}'},
	\label{GCGanzenmueller::eqn:peridynamic_energy_continuous_compact}
\end{equation}

The factor of $1/2$ in the above equation arises because each bond is defined twice, once
originating at $\vec{X}$ and pointing to $\vec{X}'$, and again via its antisymmetric counterpart
pointing from $\vec{X'}$ to $\vec{X}'$.
The forces within the bond-based PD continuum are obtained by taking the derivative of the
micropotential with respect to the bond distance vector. The \textit{microforce} between two bonded
points is thus
\begin{equation}
	\vec{f}(\vec{r}, \vec{\xi}) =
	- \frac{\partial w(\vec{r}, \vec{\xi})}{\partial \vec{r}},
	\label{GCGanzenmueller::eqn:peri_force}
\end{equation}
yielding the acceleration $\vec{a}(\vec{X})$ of a point with mass density
$\rho$ due to all its neighbors within $\mathcal{H}_{\delta}$:
\begin{equation}
	\rho \vec{a}(\vec{X}) = \int \limits_{\mathcal{H}_{\delta}} \omega(\vec{\xi}) \vec{f}(\vec{r}, \vec{\xi}) \mathrm{d} V_{\vec{X}'}.
	\label{GCGanzenmueller::eqn:peridynamic_acceleration_continuous}
\end{equation}

For implementation in a computer code, equations~(\ref{GCGanzenmueller::eqn:peridynamic_energy_continuous_compact}) and
(\ref{GCGanzenmueller::eqn:peridynamic_acceleration_continuous}) need to be discretized. This process requires the
division of the continuous body to be simulated into a number of distinct nodes with a given
subvolume, subject to the constraint that the sum of all subvolumes equals the total volume of the
body. These nodes are termed particles henceforth and the Peridynamic bonds exist between these particles.
The most straightforward discretization approach is nodal integration, which is used in almost all
publications dealing with PD up to date.
Referring to Fig.~\ref{GCGanzenmueller::fig:peridynamics_scheme}, particle $i$ is connected to all neighbors $j$
within the horizon $\delta$. Dropping the explicit dependence on $\vec{X}$, the discrete expression
for the energy density of a particle $i$ reads:
\begin{equation}
	{W}_i = \sum \limits_{j \in \mathcal{H}_{\delta}} \omega(\xi_{ij}) V_j {w}_{ij} (\vec{r}_{ij}, \vec{\xi}_{ij}),
	\label{GCGanzenmueller::eqn:peridynamic_energy_density_discretized}
\end{equation}
and 
\begin{equation}
	\vec{a}_i = \frac{1}{m_i} \sum \limits_{j \in \mathcal{H}_{\delta}} \omega(\xi_{ij}) V_i V_j \vec{f}_{ij}(\vec{r}_{ij}, \vec{\xi}_{ij}).
	\label{GCGanzenmueller::eqn:peridynamic_acc_discretized}
\end{equation}
These discretizations represent simple Riemann sums, i.e., piecewise constant approximations of
the true integrals. The object of this work is to quantify the errors incurred by this approach, but
before doing so, we introduce a specific form of the pairwise force function which is
compatible with linear elastic continuum behavior and supports a brittle fracture mechanism.


\section{Linear elasticity in Peridynamics}
In order to establish the link with linear elasticity, i.e., a Hookean solid,
Silling \cite{GCGanzenmueller::Silling:2005a} introduced the \textit{Prototype Microbrittle Material} (PMB) model, with a
microforce that depends linearly on the bond stretch $s = {|\vec{\xi} + \vec{\eta}|}/{|\vec{\xi}|}$.
The bond stretch can be thought of as a pairwise one dimensional strain description of the material,
and a full strain tensor can indeed be derived from an ensemble of bond stretches \cite{GCGanzenmueller::Silling:2007a}. A
microforce which is linear in $s$ is therefore in agreement with Hooke's law.

Here, we employ the following microforce which:
\begin{equation}
	f(s, \xi) = -c s / \xi,
	\label{GCGanzenmueller::eqn:microforce}
\end{equation}
with proportionality constant $c$. The corresponding micropotential is obtained by integrating the
microforce w.r.t. displacement.
\begin{equation}
	w(s) = - \int f(s, \xi) \mathrm{d} \eta = \frac{1}{2} c s^2.
	\label{GCGanzenmueller::eqn:micropotential}
\end{equation}
Note that the expressions for the microforce and the micropotential differ from Silling's original
work by a factor of $\xi$. This change is purely for consistency reasons, because, in our opinion, the
energy density should not contain a reference to a length scale. The modification will be absorbed
into the proportionality constant $c$ which is yet to be determined.

The weight function is chosen as a simple step function,
\begin{equation}
	\omega(\xi_{ij}) = \begin{cases} 1 & \mbox{if } \xi_{ij} \leq \delta \\
	                                      0 & \mbox{if } \xi_{ij} > \delta \end{cases},
\end{equation}
which allows for a compact notation as it can be absorbed into the summation operator of the
discretized expressions, i.e., $\sum \limits_{j \in \mathcal{H}_{\delta}} \omega(\xi_{ij}) = \sum
\limits_{j \in \mathcal{H}_{\delta}} 1$. The effects of using different weight functions have been studied
in detail \cite{GCGanzenmueller::Parks:2011a}. No
significant benefits were observed when using different forms of the weight function for the purpose
of simulating structural response problems, however, the weight function affects the dispersion of
waves.

Damage and failure are incorporated by keeping track of the history of a bond stretch state. We fail individual bonds
by permanently and irreversibly deleting them once they are stretched beyond a critical stretch
value $s_c$. 

The remaining constant $c$ is determined by requiring the Peridynamic expression for the energy
density, equation~(\ref{GCGanzenmueller::eqn:peridynamic_energy_continuous_compact}) to be consistent with the result from
linear elasticity theory, $W_{el.}$:
\begin{equation} \label{GCGanzenmueller::energy_balance_PD_continuum}
	\frac{1}{2}\int \limits_{\mathcal{H}_{\delta}} \omega(\xi_{ij}) w(s) \mathrm{d} V_{\vec{X}'} = W_{el.},
\end{equation}
In the 3D case of pure dilation or compression,  c.f.  equation~(\ref{GCGanzenmueller::eqn:Wel3D}) in the
Appendix, we have $W_{el.}^{3D} = 9 K s^2 / 2$, where $K$ is the bulk modulus and $s$ is the strain
along any of the Cartesian directions. Note that for isotropic strain field, the strain and the stretch of any bond coincide.
Integrating the Peridynamic energy density expression for this strain field in spherical coordinates, we
have
\begin{equation}
	\frac{1}{2} \int \limits_{\mathcal{H}_{\delta}} \omega(\xi_{ij}) w(s) \mathrm{d} V
	= \frac{1}{2} \int \limits_{0}^{\delta} \int \limits_{0}^{\pi} \int \limits_{0}^{2 \pi}
          \omega(\xi_{ij}) \frac{1}{2} c s^2 \xi^2 \mathrm{d}\xi \sin(\phi) \mathrm{d} \phi \mathrm{d} \theta
	= \frac{\pi c s^2 \delta^3} {6}.
	\label{GCGanzenmueller::eqn:W_PD_integral}
\end{equation}
Equating this result with the continuum theory expression for the elastic strain energy, the
constant $c$ is obtained as:
\begin{equation}
	c = \frac{6 K} {\pi \delta^3}.
	\label{GCGanzenmueller::eqn:c_integral}
\end{equation}
This approach of determining $c$ is correct for the continuous integral expressions upon which PD
theory is based. However, in combination with the discrete expression given by
equation~(\ref{GCGanzenmueller::eqn:peridynamic_energy_density_discretized}), the results of a numerical computation of
the energy density are inaccurate, as exact analytic integration is combined with piecewise constant
approximation of the integrals.  The errors incurred by this approach are rather large and, what is
worse, does not converge to zero upon increasing $\delta$ or the number of particles. Before we
quantify these errors, we introduce an alternative approach to determine $c$ which relies on exact
error cancellation such that the energy density is exactly reproduced for a given strain field.


\subsection{An improved route for determining the PMB proportionality constant}

\begin{figure}[!ht]
\begin{center}
\includegraphics[width=4in]{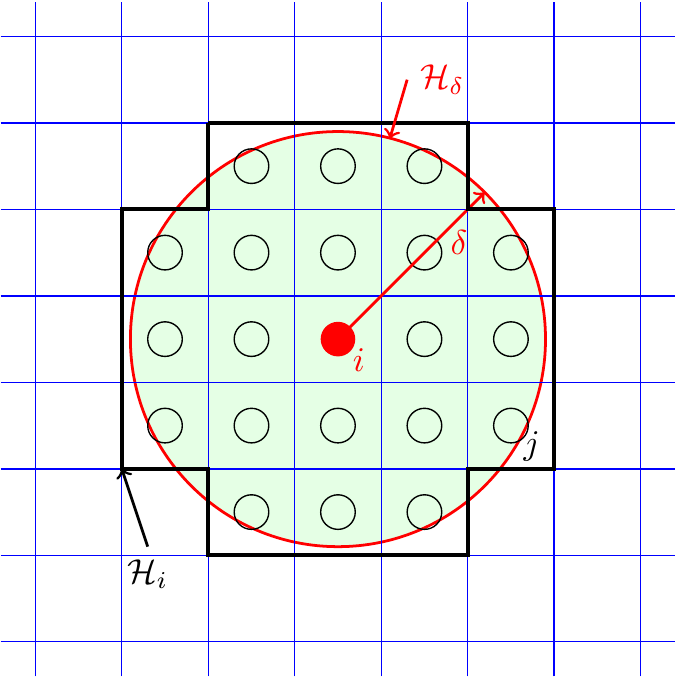}
\end{center}
\caption{
Piecewise constant approximation of the Peridynamic neighborhood volume.
In the original derivation of Peridynamics, the neighborhood boundary $\mathcal{H}_{\delta}$ is
defined as a smooth region in space given by the radial cutoff $\delta$.
For a piecewise constant approximation of the integrals of the Peridynamic
theory suitable for computer implementation, the neighborhood also needs to be defined in a discrete
manner: here, we define the piecewise constant neighborhood approximation as the volume of all
particles touched by the radial cutoff $\delta$.}
\label{GCGanzenmueller::fig:PD_neighborhood_piecewise}
\end{figure}

Instead of deriving the proportionality constant $c$ by exact analytic integration, we propose to
use the same integral approximation as is used for discretizing the PD energy density integral or
acceleration expression. This means that we use a piecewise constant approximation for
equation~(\ref{GCGanzenmueller::energy_balance_PD_continuum}), as shown in Fig.~\ref{GCGanzenmueller::fig:PD_neighborhood_piecewise}:
\begin{equation}
	\frac{1}{2}\int \limits_{\mathcal{H}_{\delta}}  w(s) \mathrm{d} V_{\vec{X}'} \approx
        \frac{1}{2} \sum \limits_{j \in \mathcal{H}_i} w(s) V_j = W_{el.}
\end{equation} 
Inserting the micropotential and the 3D pure dilation result for the continuum strain energy
density in the above equation, we obtain the proportionality constant as
\begin{equation}
	c_i = \frac{18 K} {\sum \limits_{j \in \mathcal{H}_i} V_j}.
	\label{GCGanzenmueller::eqn:c_piecewise_3D}
\end{equation}
In this formulation, the dependence of $c_i$ on the horizon $\delta$ is now only implicit through the
number of particles contributing to the sum in the denominator.  A particle at a free surface of a
body will have a different number of neighbors compared to a particle in the bulk. This effect is
accounted for with our discrete expression for $c_i$, as opposed to the original expression,
equation~(\ref{GCGanzenmueller::eqn:c_integral}), which is only valid for the bulk. This normalization is similar to a
Shepard correction of the shape functions encountered in other meshless methods such as
Smooth-Particle Hydrodynamics \cite{GCGanzenmueller::Shepard:1968a, GCGanzenmueller::Randles:1996a}, where it restores $C^0$ consistency, i.e., the
ability to approximate a constant field. At the same time, it is this local dependence which allows
us to easily introduce different spatial resolutions and horizons. It is important at this point to
discuss the conservation of momentum. In the original formulation of the PMB model, the
proportionality constant $c$ is the same for all interacting particles. Therefore, $\vec{f}_{ij} = -
\vec{f}_{ji}$, and, as the forces are aligned with the distance vector between particles $i$ and $j$,
both linear and angular momentum are conserved. In the approach proposed here, $\vec{f}_{ij}$ is not
necessarily equal to $-\vec{f}_{ji}$, as the particle volume sum over $\mathcal{H}_i$ is not
guaranteed to equal the particle volumes sum over $\mathcal{H}_j$. Thus $c_i \neq c_j$, in
general. We therefore enforce symmetry in the following manner:
\begin{equation}
	c_{ij} = \frac{c_i + c_j}{2}
	\label{GCGanzenmueller::eqn:c_symmetric}
\end{equation}
The full expressions for the potential energy of a particle and its acceleration, as required for
implementation in a computer code, are then
\begin{equation}
	{E}_i = \sum \limits_{j \in \mathcal{H}_i} V_i V_j c_{ij} s_{ij}^2,
	\label{GCGanzenmueller::eqn:peridynamic_energy_gcg}
\end{equation}
and 
\begin{equation}
	\vec{a}_i = \frac{1}{m_i} \sum \limits_{j \in \mathcal{H}_i} V_i V_j c_{ij} s_{ij}
\frac{1}{\xi_{ij}} \frac{\vec{r}_{ij}}{r_{ij}}.
	\label{GCGanzenmueller::eqn:peridynamic_acc_gcg}
\end{equation}


\section{Results}
\subsection{Comparison of the original PMB model with the improved model}
This section presents two examples to assess the accuracy of the original PMB model and the
normalization procedure proposed in this work. We show that the energy density and speed of
sound are exactly reproduced using our method, while the original method yields considerable errors.
Finally, we investigate a mode-I crack opening example with our modified PD scheme, where a failure
criterion based on the Griffith energy release rate correctly reproduces results from Linear
Elasticity Fracture Mechanics Theory.

\subsubsection{Energy density}
\begin{figure}[!ht]
\begin{center}
\includegraphics[width=4in]{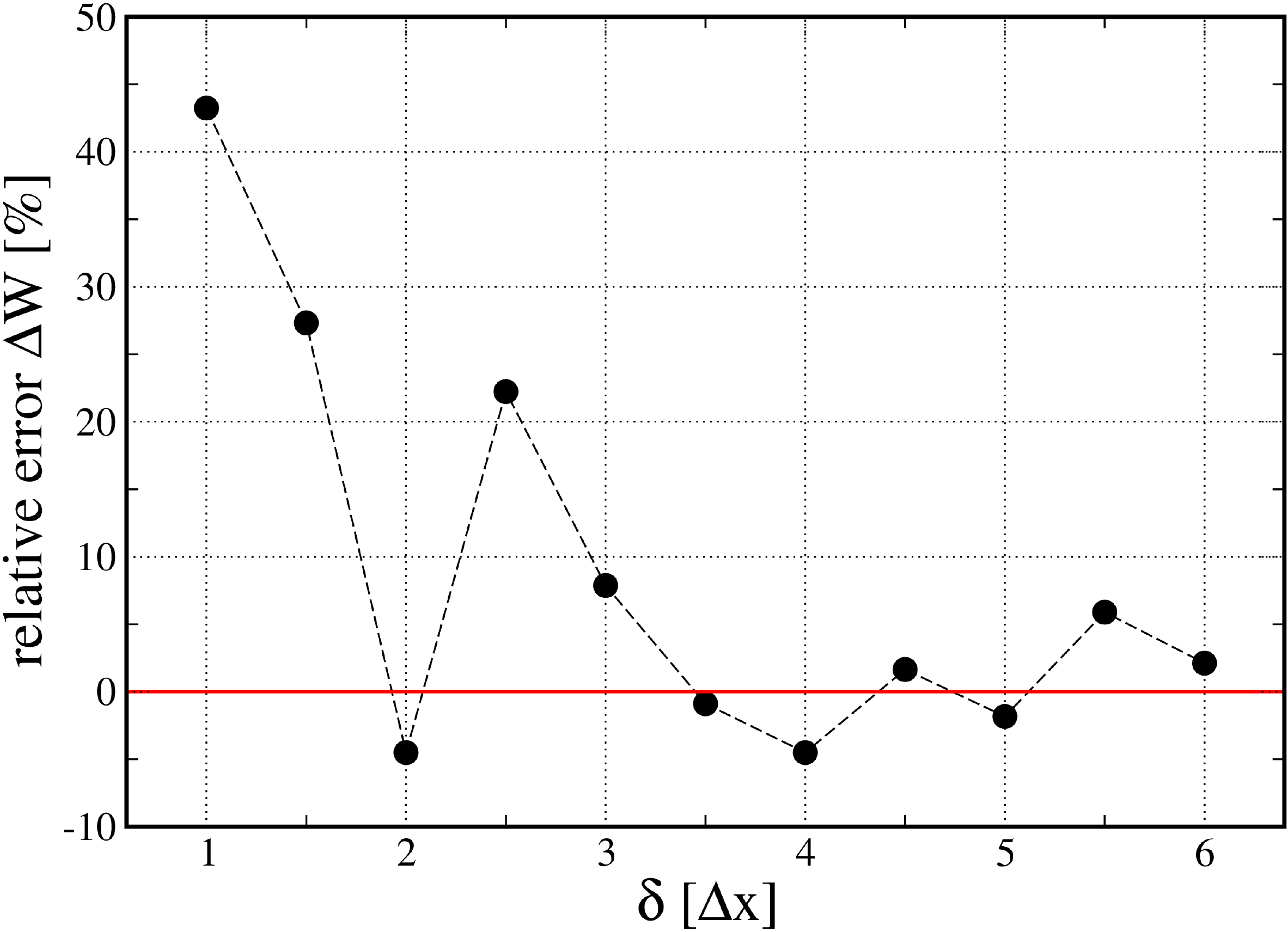}
\end{center}
\caption{
This graph shows relative errors of the Peridynamic strain energy density for a pure dilation
strain field. Black symbols denote results obtained with the original PMB method which uses analytic
integration for the determination of the micropotential proportionality constant. The red line shows the
results obtained using the here proposed normalization approach for the micropotential constant.}
\label{GCGanzenmueller::fig:strain_energy_error}
\end{figure}

The ability to reproduce the correct strain energy for a homogeneous deformation
is the most basic task any simulation method for solid mechanics should be able to
handle with good accuracy.  We consider a cube of a material under periodic
boundary conditions. The bulk modulus is 1 GPa, and the material is discretized
using a cubic lattice with spacing $\Delta x = 1$ m. In order to effect a
homogeneous deformation, all directions are scaled using a factor of
$l=\mathrm{1.05}$, leading to volume change of 15.8\%.  We measure the
Peridynamic strain energy density, $W_{PD}$ by summing over all bond energies
and dividing by the cube volume. The exact strain energy density is calculated
using equation~(\ref{GCGanzenmueller::eqn:Wel3D}), such that a relative error can be defined:
\begin{equation}
	\Delta W = \frac{W_{PD} - W_{el.}^{3D}}{W_{el.}^{3D}}\:.
\end{equation}
Fig.~\ref{GCGanzenmueller::fig:strain_energy_error} shows the relative errors for the original
method and a range of different horizon cutoffs $\delta \in [2 \Delta x \ldots
6 \Delta x]$, such that the number of particles within the horizon varies from
32 to 924.  We observe that the original approach shows relative errors in excess of
30\%. What is worse, is that the errors do not converge monotonously as one
increases the horizon, which is the only resolution variable available due to
the scale invariance implied by the absence of free surfaces. In contrast, the
normalization proposed here reproduces the strain energy density exactly,
within numerical precision.

\subsubsection{Wave propagation}

\begin{figure}[!ht]
\begin{center}
\includegraphics[width=4in]{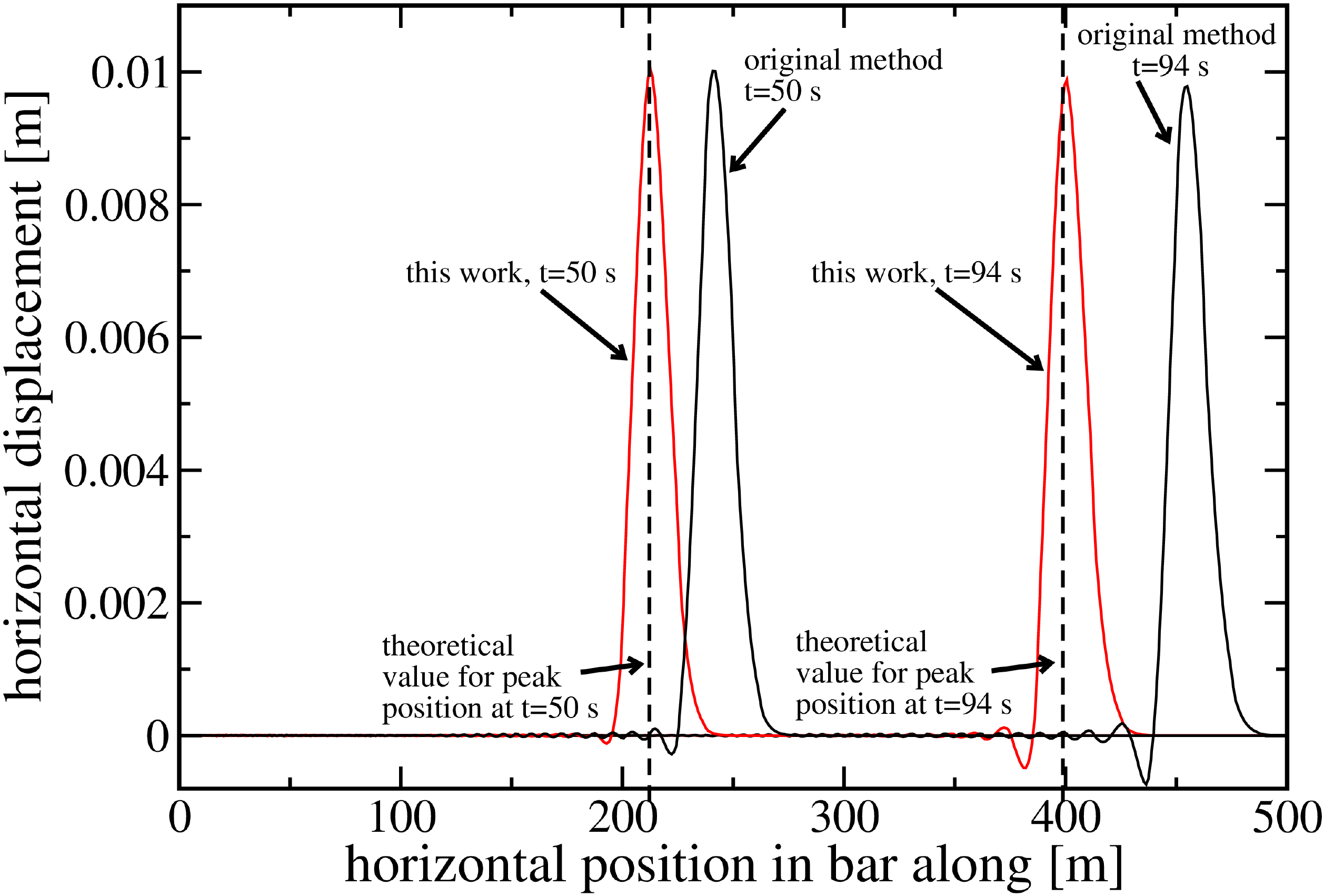}
\end{center}
\caption{
Sound wave propagation. A horizontally oriented bar of dimensions 500 m * 4 m * 4 m is loaded using a Gaussian shaped displacement at the left. This initial perturbation causes a Gaussian-shaped pressure pulse to travel to the right at the longitudinal speed of sound. Shown above are theoretical values (vertical dashed lines), where the center of the pressure pulse should be located after elapsed time periods of 50 s and 94 seconds, respectively. The results of the Peridynamics simulations, (i) using the original approach, and (ii) using the normalization for the micropotential amplitude are shown as black and red lines, respectively. We note that the original approach over-predicts the speed of sound by 13\%, while the here proposed normalization approach agrees with the theoretical result within an error margin of less than 1\%.
}
\label{GCGanzenmueller::fig:sound_wave}
\end{figure}

The second example investigates the propagation of a pressure pulse. To this end, we consider a
bar of size $500\times4\times4\:\mathrm{m}^3$, discretized using a cubic lattice with $\Delta
x=1\:\mathrm{m}$. We set $K=1\:\mathrm{Pa}$, $\rho=1\:\mathrm{kg/m}^3$ and $\delta=2.5
\Delta x$.  Periodic boundaries are applied along the $y$- and $z$-direction in order to
suppress free surface effects. The pulse is initiated by a displacement perturbation of Gaussian
shape at one end,
\begin{equation}
\vec{x} = \vec{X} + 0.02\;\mathrm{m} \: \times\: \exp\left(-\frac{\vec{X} \cdot \vec{X}}{100\;\mathrm{m}^2}\right) \vec{e}_x,
\end{equation}
where $\vec{e}_x$ is the unit vector in the Cartesian x-direction.
The simulation is then run until the pressure pulse has reached the right end of the bar. The
time-step is set to $\Delta t = 0.1\:\mathrm{s}$, which is stable according to CFL
analysis. Following \cite{GCGanzenmueller::Kinsler:2000}, the theoretical value for the longitudinal speed of sound is 
\begin{equation}
	c_l = \sqrt{\frac{K + \frac{3}{4} G}{\rho}},
\end{equation}
where $G=3K(1-2\nu)/[2(1+\nu)]$ is the shear modulus, and $\nu$ is Poisson's ratio. As the 3D
Peridynamic model under consideration has a fixed Poisson ratio $\nu=1/4$ \cite{GCGanzenmueller::Silling:2000a}, we obtain $c_l=4.24\:\mathrm{m/s}$.
Fig.~\ref{GCGanzenmueller::fig:sound_wave} compares this theoretical prediction with the results of
Peridynamics simulation that employ the original analytical integration approach for determining the
amplitude constant $c$ of the micropotential, and the normalization approach proposed here. It is
evident from this comparison that the original approach severely overestimates the wave propagation
speed. This is in agreement with the observation, that the original approach overestimates the
energy density, leading to a system which is effectively too stiff. In contrast, the normalization
procedure for determining $c$ reproduces the theoretical wave propagation speed very well.

\begin{figure}[!ht]
\begin{center}
\includegraphics[width=4in]{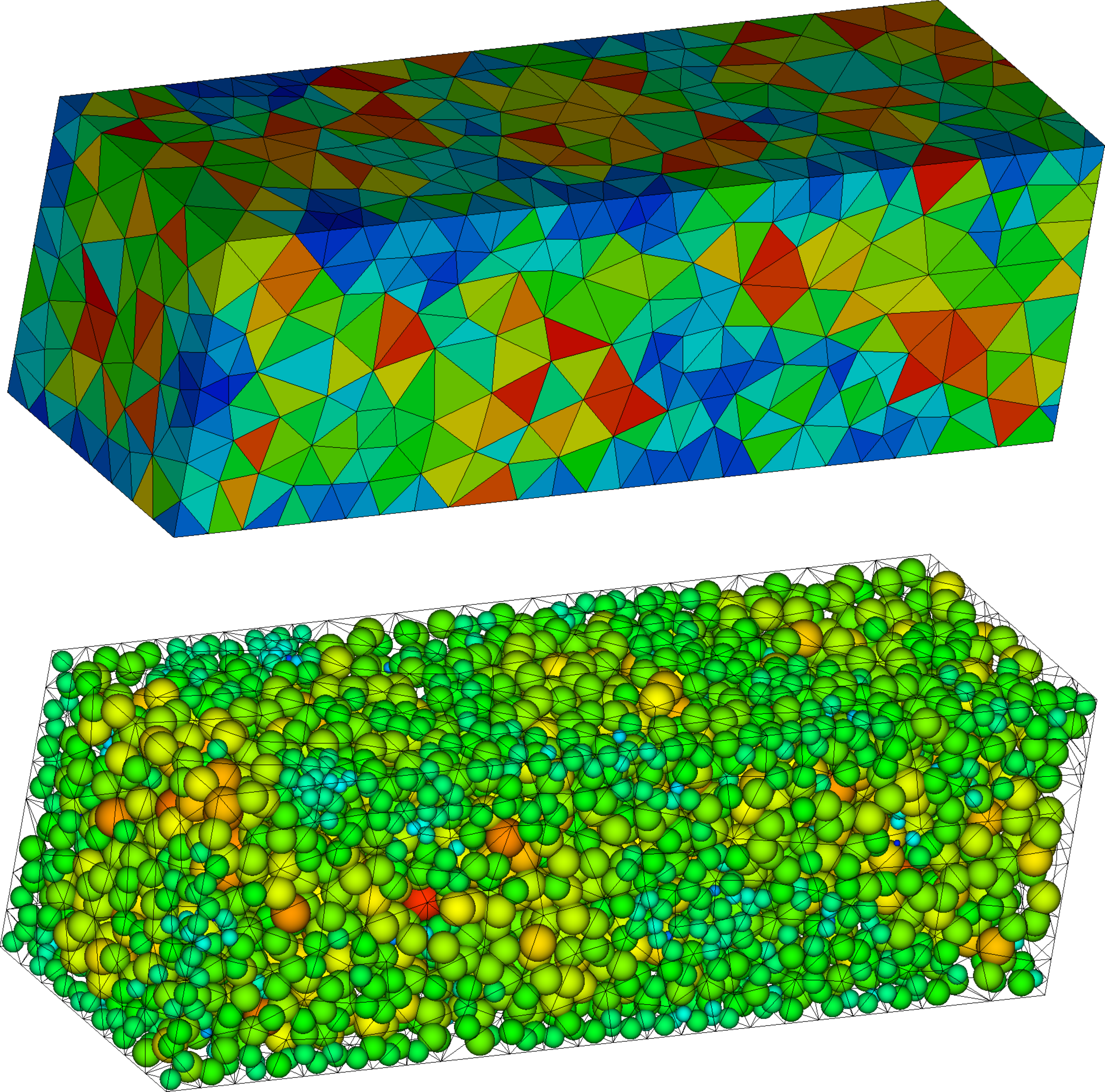}
\end{center}
\caption{
Generation of non-uniform particle configurations. Top: a volume is meshed using regular
tetrahedrons. Form this mesh, a particle configuration is obtained by placing particles at the
tetrahedron barycenter, and assigning the tetrahedron's volume and mass to the particles. Color coding
represents volume, increasing from blue to red.
}
\label{GCGanzenmueller::fig:tet_mesh}
\end{figure}

To investigate the performance of the normalization approach in the case of non-uniform particle spacing,
we now consider a mesh of the same bar as above, which is generated via a stochastic procedure. We
use a Delauney-based meshing algorithm to generate tetrahedral elements. These elements are
subsequently replaced by particles. Each particle is assigned the volume of the tetrahedron it
replaces. The particle's mass is obtained from the volume and the mass density, $m=V\rho$.
Fig.~\ref{GCGanzenmueller::fig:tet_mesh} shows a section of the bar in both the tetrahedron and particle
representation. To realize a challenging test, the tetrahedral mesh was intentionally generated such
that small angles and large variations in the tetrahedron volumes are achieved. The resulting
particle configuration is therefore strongly polydisperse with a ratio of smallest to largest radius
of 100.  Because no characteristic length-scale (such as the lattice spacing above) is now present,
we adjust the Peridynamic horizon for each particle separately, such that the neighborhood contains
30 neighbors. Three different initial tetrahedron meshes of different resolutions are used to
conduct a convergence study for our PMB normalization approach. The coarsest mesh contains 17211
tetrahedrons, and two more finely resolved meshes are obtained by repeated splitting of the elements,
such that the finest mesh has 70381 elements.

\begin{figure}[!ht]
\begin{center}
\includegraphics[width=4in]{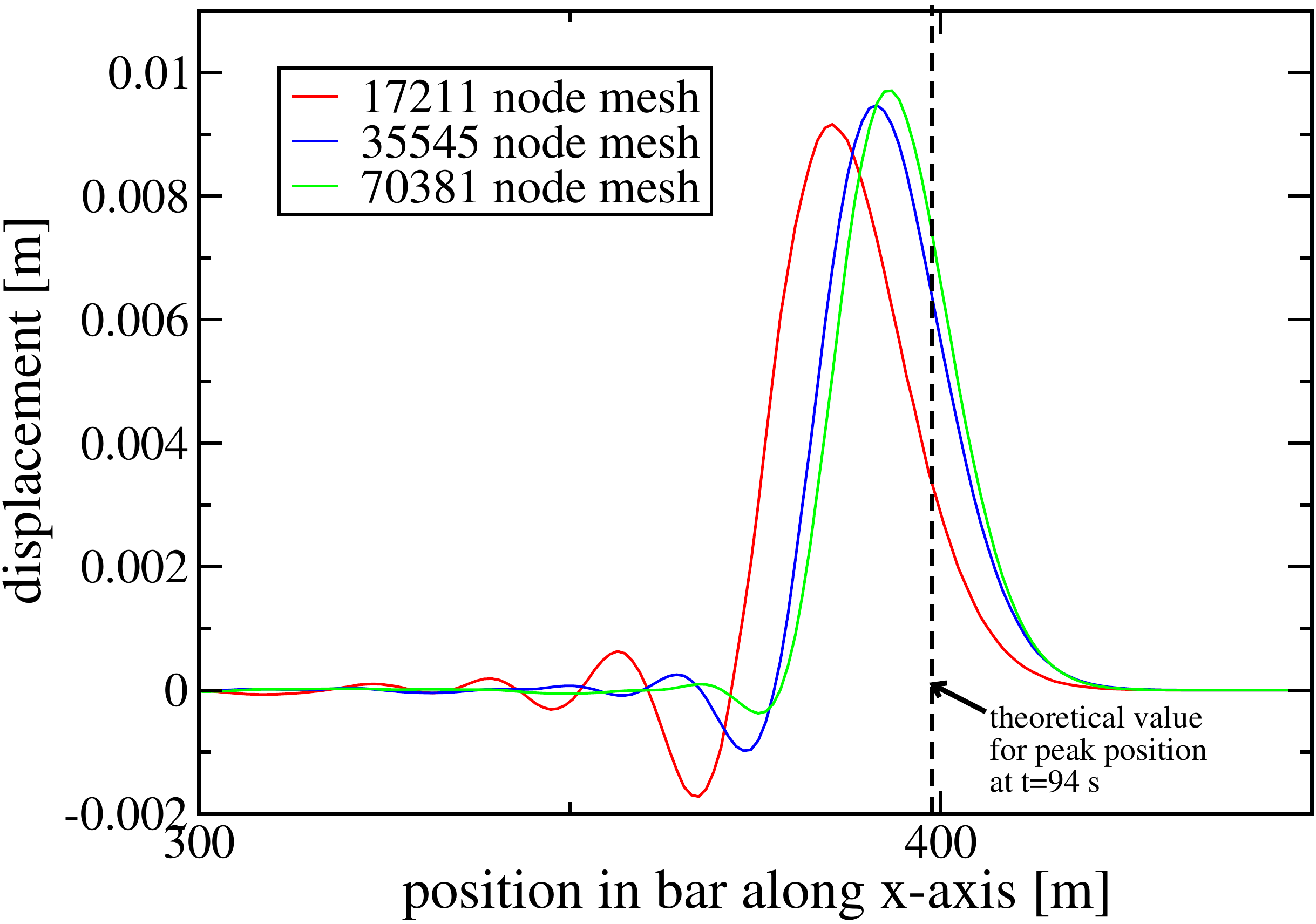}
\end{center}
\caption{
Propagation of a pressure pulse in a long bar which is discretized using irregular particle positions
and polydisperse particle size distributions. The geometry and parameters are the same as for 
Fig.~\ref{GCGanzenmueller::fig:sound_wave}, but instead of a regular mesh we employ the discretization
approach via a stochastic tetrahedral mesh outlined in Fig.~\ref{GCGanzenmueller::fig:tet_mesh}.
The vertical dashed line indicates the position where the pressure pulse should be, according to the
exact wave propagation speed. The Peridynamic simulations show convergence to this exact result 
upon increasing the number of particles used for discretizing the bar.}
\label{GCGanzenmueller::fig:sound_wave_tet_mesh}
\end{figure}

The results are given in
Fig.~\ref{GCGanzenmueller::fig:sound_wave_tet_mesh}. We observe that pressure pulse is much
broader when compared to the results of the uniform particle configuration shown in 
Fig.~\ref{GCGanzenmueller::fig:sound_wave}, and that oscillations travelling behind the main pulse are
more pronounced. This is not surprising, as it is well known that wave propagation is affected by
discretization effects: partial reflections occur always when a wave is transmitted between regions
of space that are discretized using different resolutions. These reflections cause dispersion and
reduction in the observed wave speed propagation speed. As the discretization length scale becomes
small compared to the wavelength, these effects disappear. We therefore expect convergence of the
location of the pressure pulse to its theoretical position at a given time, and return of its shape
back to the initial Gaussian shape, as the particles are more finely resolved. The simulation
results shown in Fig.~\ref{GCGanzenmueller::fig:sound_wave_tet_mesh} support these statements: as
the resolution is enhanced, the wave speed tends towards its theoretical value and the pressure
pulse shows less oscillations. We therefore conclude that our approach of handling interactions
between Peridynamic particles of different size is correct. 

\subsection{Fracture energy}

Traditionally, continuum mechanics is formulated using a set of partial differential equations which
describe temporal and spatial evolution. These equations require smooth solutions with well defined
gradients. Therefore, discontinuities in the material, such as cracks, cannot emerge naturally
within the solution manifold. In contrast, Peridynamics circumvents this problem by employing an
integral description for the evolution equations. Due to its simple form, the PMB model in
particular is well suited to model arbitrary crack initiation and propagation phenomena.  A number
of studies have used the PMB model to study crack propagation speed, crack branching as well as
coalescence of individual cracks \cite{GCGanzenmueller::Silling:2010a, GCGanzenmueller::Ha:2010a,
GCGanzenmueller::Ha:2011, GCGanzenmueller::Agwai:2011a}. However, to the best of these authors'
knowledge, no quantitative assessment of the accuracy of PMB simulations relative to analytical
solutions for modelling crack initiation and propagation has been published to date. The main reason
for this shortcoming is probably the fact that the original formulation of the PMB model using the
analytic integration approach for determining the micropotential amplitude inflicts unacceptably
large errors already for the energy density.  This implies that no quantitatively correct modelling
of crack processes could be carried using the original PMB approach. However, the above cited
studies demonstrate that the original PMB model is very well suited to qualitatively model complex
crack growth phenomena, including the interaction of multiple cracks with each other.  In this
section, we demonstrate the our normalization approach for determining the micropotential amplitude
can be used to quantitatively reproduce analytical solutions obtained from Linear Elastic Fracture
Mechanics (LEFM) Theory.

\begin{figure}[!ht]
\begin{center}
\includegraphics[width=0.75 \textwidth]{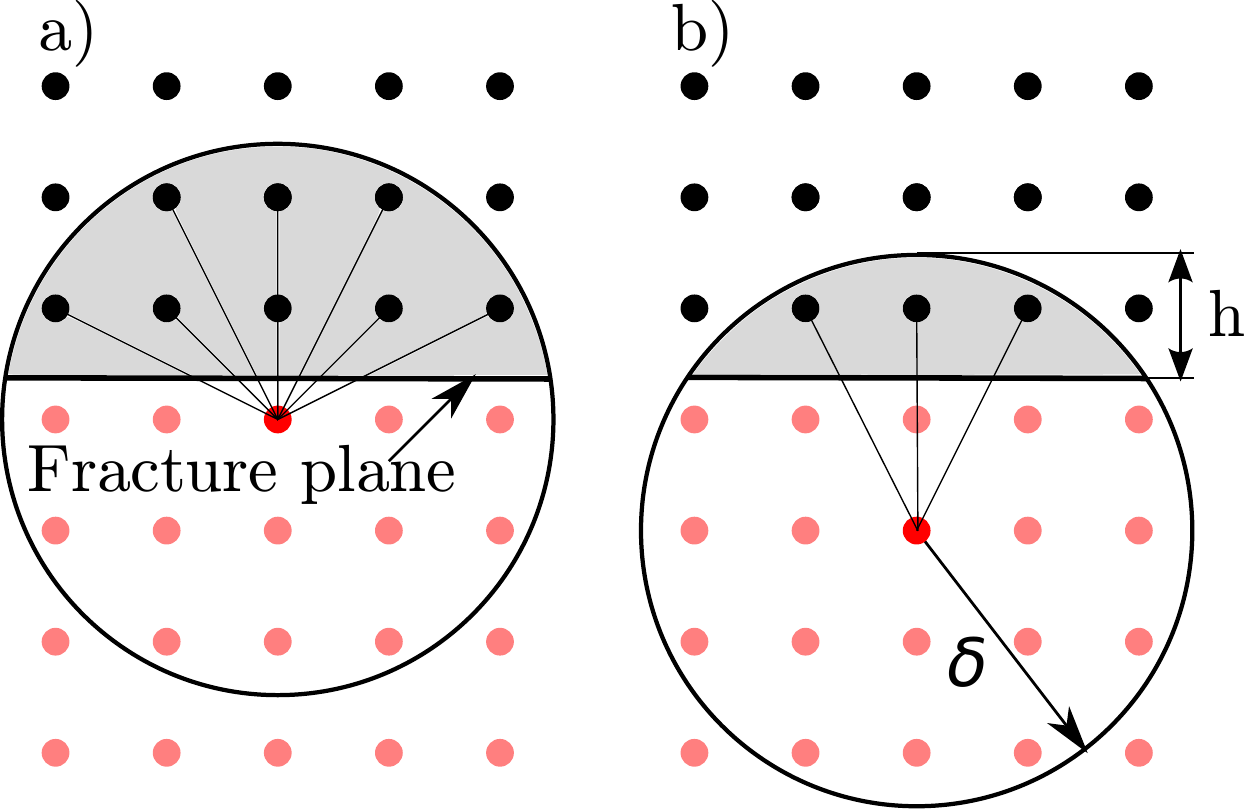}
\end{center}
\caption{
Peridynamic bond interactions across a hypothetical fracture surface in a 2D plane strain
model. Configurations a) and b) show two examples for bonds which traverse a hypothetical fracture
surface.  The total energy which is set free if the hypothetical fracture surface becomes real is
the sum of the energies stored in all the bonds between the red and black particle half-spaces.
While it is in principle possible to enumerate these bonds and perform an explicit summation, this
approach is cumbersome in practice. Instead, we consider the interaction volume to either side of
the fracture surface, which is given by the plane thickness multiplied with the circular segment (gray),
$V_c(h, \delta) = t {\delta}^{2}\arccos \left( {\frac {\delta-h}{\delta}} \right) - \left( \delta-h
\right) \sqrt {2\,\delta h-{h}^{2}}$. The energy density of each configuration is given by the
product of the the micropotential and the interaction volume. Finally, the energy release rate is
obtained by integrating the energy density over all configurations by varying $h$.}
\label{GCGanzenmueller::fig:2D_crack_surface}
\end{figure}

A useful crack propagation theory for numerical simulations must be based on criteria which are
independent of the discretization length scale. If length scale-dependent measures such as stress
are used instead, no convergence of the loads required to propagate a crack can be achieved because
finer resolution always implies a higher stress concentrations. One useful criterion is the Griffith energy release
rate, i.e., the energy required to separate a body by generating two free surfaces, one to either
side of a crack area. The energy release rate is defined as energy divided by area and is therefore
an intensive measure for the resistance of a body against cracking. In the discrete setting of a
numerical simulation, the energy release rate incorporates the discretization length scale and thus
provides a failure criterion which is independent of discretization. This implies that a crack growth
simulation based on such a failure criterion can converge upon discretization refinement.
A Peridynamic failure criterion based on the Griffith energy release rate has been first published
by Silling and Askari \cite{GCGanzenmueller::Silling:2005a}. Here, we roughly follow their approach,
but restrict ourselves to plane-strain conditions as LEFM Theory provides useful analytical
solutions to compare against in this case. 

\begin{figure}[!ht]
\begin{center}
\includegraphics[width=0.3 \textwidth]{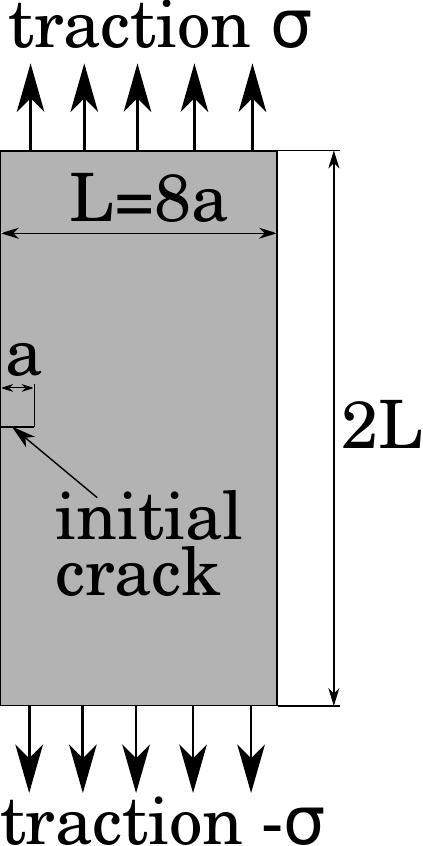}
\end{center}
\caption{
Sketch of the geometry used for the crack propagation analysis. A rectangular patch of a
linear-elastic material is stretched by applying tractions to the top and bottom side. The patch
features an initial crack which serves to effect stress concentration at the crack tip. This
geometry and loading scenario can be solved analytically solved for a critical traction which causes
the crack to grow using Linear Elastic Fracture Mechanic Theory.}
\label{GCGanzenmueller::fig:2D_crack_sketch}
\end{figure}

\begin{figure}[!ht]
\begin{center}
\includegraphics[width=0.9 \textwidth]{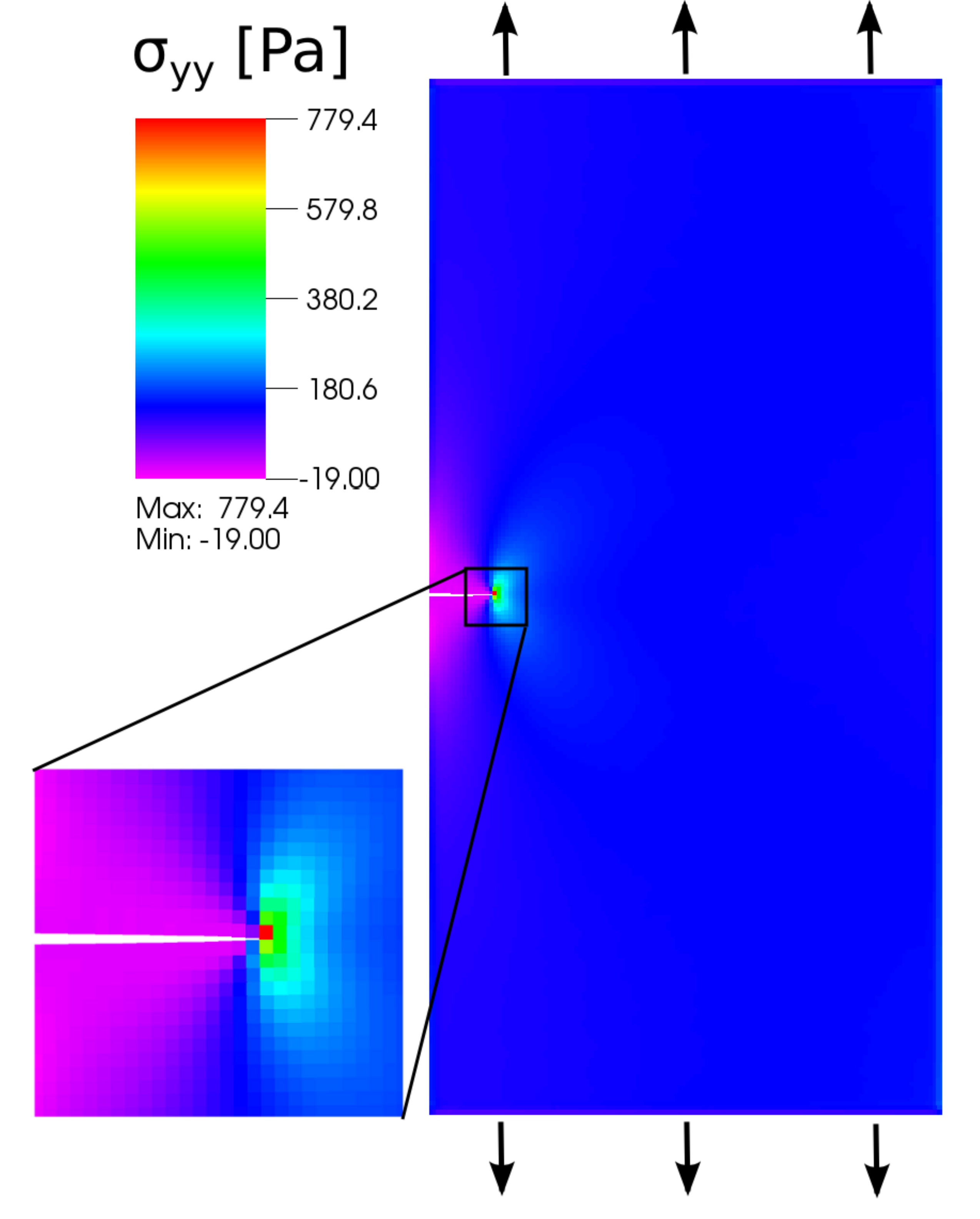}
\end{center}
\caption{
Peridynamic simulation of crack propagation. Shown is a snapshot of the simulation with the finest
resolution. The color-coding represents the yy-component of the stress tensor. The zoomed-in area
shows the stress concentration at the crack tip.}
\label{GCGanzenmueller::fig:2D_crack_snapshot}
\end{figure}

\begin{figure}[!ht]
\begin{center}
\includegraphics[width=0.75 \textwidth]{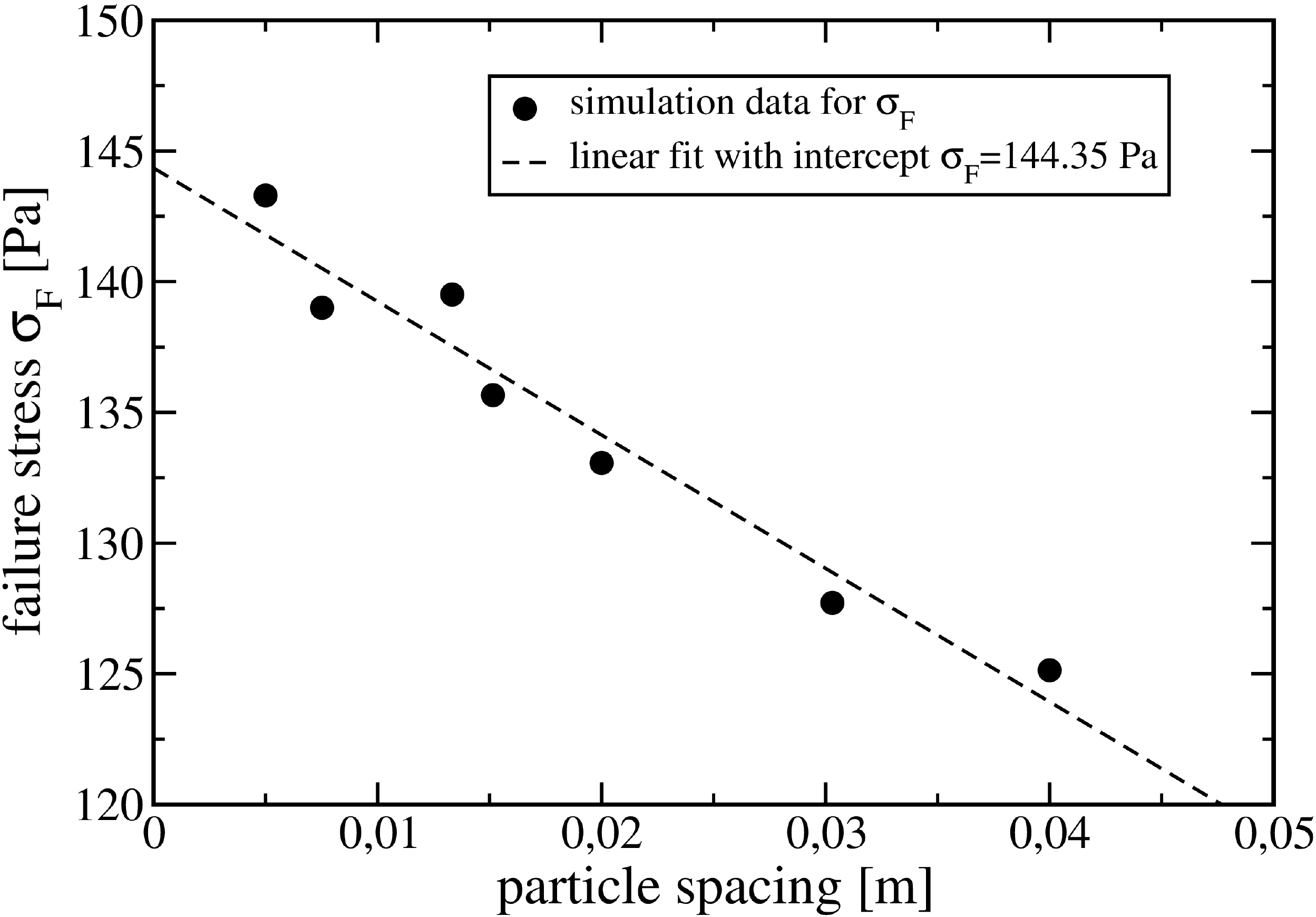}
\end{center}
\caption{
Convergence of the critical tractions required to cause abrupt crack propagation. As the particle
spacing is reduced, linear convergence towards the exact result $\sigma_F=146.9\;\mathrm{PA}$ is
observed.
}
\label{GCGanzenmueller::fig:2D_crack_convergence}
\end{figure}

Because PMB interactions are formulated in terms of bond-wise micropotentials, a failure
criterion is required which links the micropotential to the energy release rate. Such an expression
can be obtained by considering a pure dilation stretch state of a Peridynamic material and summing
the energy stored in all those bonds which cross a hypothetical unit fracture surface. The resulting
normalized energy per area, which is a function of the bond stretch and the bulk modulus, can be
equated with the energy release rate. From this relation a critical bond stretch can be obtained at
which the bond should fail in order to yield a given energy release rate.
Fig.~\ref{GCGanzenmueller::fig:2D_crack_surface} shows how Peridynamic bonds which are connected to a particular central
node interact across a hypothetical fracture surface. An interaction volume is defined as as the
spatial volume occupied by these bonds. For a given fracture surface, a manifold of interaction volumes exist. The magnitude of these
volumes depends on the distance of the central node away from the fracture surface. Thus, we obtain
the Peridynamic energy release rate, $G_{I,PD}$, by integrating the product of micropotential and
interaction volume over all values of the distance of the central node to the fracture surface.
Referring to Fig.~\ref{GCGanzenmueller::fig:2D_crack_surface}, this integral is given by:
\begin{eqnarray} 
	G_{I,PD} & = & 2\; \int\limits_{h=0}^{\delta} w(s)\; V_c(h,\delta) \mathrm{d}h \nonumber \\
                 & = & 2\; \int\limits_{h=0}^{\delta} \left[ \frac{1}{2} c s^2 \; t {\delta}^{2}\arccos \left(
{\frac {\delta-h}{\delta}} \right) - \left( \delta-h \right) \sqrt{2\,\delta h-{h}^{2}} \right] \mathrm{d}h \nonumber \\
                 & = & \frac{2}{3} c s^2 t \delta^3.
\end{eqnarray}
Note that the factor of 2 in front of the integral stems from the fact that we have
two interaction volumes, one to either side of the hypothetical fracture surface. The factor $t$
above is the thickness of the plane-strain model.
Requiring that the Peridynamic energy release rate matches a specified energy release rate,
$G_{I,PD} = G_I$ we obtain the critical bond stretch at failure as:
\begin{equation}
	s_c = \sqrt{\frac{3 G_I}{2 c t \delta^3}}.
\end{equation}
A useful test for the above expression is delivered by LEFM Theory, which provides analytical
solutions that predict the onset of crack growth for some simple models. One such model
is a  rectangular patch of an elastic material with an existing sharp
crack on one side, which is stretched by applying tractions, see Fig.~\ref{GCGanzenmueller::fig:2D_crack_sketch}.
For prescribed values of the energy release rate and the Young's modulus, a critical traction is
predicted by LEFM Theory when failure should occur by abrupt propagation of the initial crack
through the entire patch. For this geometry, the critical traction that leads to
failure is known to be \cite{GCGanzenmueller::Cook:1999}
\begin{equation}
\sigma_F = {\it K_I}{\frac {1}{\sqrt {\pi \,a}}} \left(  1.12- 0.23\,{\frac {a}{L}
}+ 10.6\,{\frac {{a}^{2}}{{L}^{2}}}- 21.7\,{\frac {{a}^{3}}{{L}^{3}}}+
 30.4\,{\frac {{a}^{4}}{{L}^{4}}} \right) ^{-1}.
\end{equation}
Here, $\sigma_F$ is the traction applied to the top and bottom of the patch which causes the crack
to propagate, $a$ is the initial length of the crack, $L$ is the width of the patch, and $K_I$ is
the fracture toughness. In plane strain, the fracture toughness can calculated from the Griffith
energy release rate $G_I$, the Young's modulus of the system, and the Poisson ratio:
\begin{equation}
	K_I = \sqrt{\frac{G_I\,E}{1-\nu^2}}
\end{equation}
With the values $E=10^4$ Pa, $\nu=1/3$, $G_I=1\;\mathrm{J}/\mathrm{m}^2$, $L=1\;\mathrm{m}$ and
$a=L/8$, we obtain the failure traction as $\sigma_F = 146.9\;\mathrm{Pa}$. This result will serve
as the reference solution against which the normalized PMB model presented in this work will be
compared.
Peridynamic simulations were carried out using a square lattice discretization of this geometry with
seven different lattice constants ranging from 0.005 m to 0.04 m, resulting in total particle
numbers from 1,250 to 80,000. Tractions were realized by gradually applying opposite forces to the
top and bottom row of particles, effecting a gradual stretch of the patch. The forces were ramped up
in time such that a displacement velocity $10^4$ times slower than the speed of sound in the patch
was achieved. Under these conditions, the simulation can be effectively considered quasi-static.
Fig.~\ref{GCGanzenmueller::fig:2D_crack_snapshot} shows a snapshot of the simulation with the highest resolution,
just before the crack starts to grow. In Fig.~\ref{GCGanzenmueller::fig:2D_crack_convergence}, the traction values
are reported for each resolution, when the crack starts to grow. These data points suggest linear
convergence of the critical traction towards the analytical result from above: the extrapolated
infinite-resolution simulation value is $144.4\pm1.5$ Pa, while the analytical result is 146.9 Pa.
The agreement between these results is very good and we attribute the remaining difference to the fact
that the initial crack does not, depending on the actual particle spacing, align perfectly with the
particles. This observation can also explain the scattering of the data points around the linear
fit, because, the simulated initial crack is sometimes shorter or longer by one lattice constant
when compared to what it should be. Nevertheless, we note that the simple normalized PMB model
is highly successful at predicting the correct stress at the crack tip which causes the crack to
grow.

\section{Discussion}
We have shown that the discrete implementations of the original formulation of the
Prototype-Microbrittle Model of linear elasticity in Peridynamics suffers from severe inaccuracies.
The origin of this deficiency is traced back to the way  how the micropotential proportionality
constant is derived. The original approach employs exact analytical integration for this quantity.
In a numerical implementation, however, field variables depending on the micropotential are
evaluated using non-exact integration rule, e.g., piecewise constant integration via the Riemann sum.
The inconsistency between these different integration approaches causes inaccuracies. To resolve
this problem, we have modified the PMB model such that the same numerical integration rule is used
for determining both the micropotential proportionality constant and the field variables. As an
additional result, interactions between particles with different sizes and different Peridynamic
horizons can be natively treated using our modification. The correctness of the new approach is
validated by simulating the propagation of sound waves, where very good agreement with the
theoretical prediction is observed. It is instructive to interpret our modification as a
normalization procedure, which performs so well because it effects error cancellation. The modified
PMB scheme bears strong similarity to other meshless simulation methods such as Smooth-Particle
Hydrodynamics, where such a normalization is known as the Shepard correction. Because Peridynamics
is most useful for dealing with material discontinuities, we also consider a crack initiation and
propagation example. Here, a patch of an elastic material with a pre-existing crack is pulled apart.
Once a critical traction is reached, the stress concentration at the existing crack tip cause the
crack to grow abruptly and cause complete separation of the patch. Peridynamic simulations of this
experiment with the modified PMB model show linear convergence to the exact critical traction as the
discretization resolution is enhanced. Much praise has been granted in advance to Peridynamics as a
method specifically apt to handle complex crack growth phenomena. The simulations reported herein
constitute the the first quantitative demonstration that Peridynamics is indeed able to correctly
predict failure in agreement with exact analytical solutions.


\section{Appendix}
\subsection{Strain energy density}
In the continuum theory of linear elasticity, the stress tensor $\vec{\sigma}$ is obtained from a
linear relationship between the stiffness tensor $\vec{C}$ and the strain tensor $\vec{\epsilon}$,
\begin{equation}
	\sigma_{ij} = C_{ijkl} \epsilon_{kl}.
\end{equation}
Employing Voigt notation \cite{GCGanzenmueller::VoigtNotation} to reduce the dimensionality of the
above tensors, the stiffness tensor is expressed as a 6x6 matrix in terms of bulk modulus $K$ and
Poisson's ratio $\nu$ as,
\begin{equation}
\vec{C} = \frac{3 K}{1 + \nu}
\left[ \begin {array}{cccccc} 1-\nu&\nu&\nu&0&0&0
\\ \noalign{\medskip}\nu&1-\nu&\nu&0&0&0\\ \noalign{\medskip}\nu&\nu&1
-\nu&0&0&0\\ \noalign{\medskip}0&0&0&1/2-\nu&0&0\\ \noalign{\medskip}0
&0&0&0&1/2-\nu&0\\ \noalign{\medskip}0&0&0&0&0&1/2-\nu\end {array}
 \right],
\end{equation}
and the symmetric stress and strain tensors reduce to vectors with six entries:
\begin{equation}
\vec{\epsilon} = 
\left[ \begin {array}{c} \epsilon_{xx}\\ \noalign{\medskip} \epsilon_{yy}\\ \noalign{\medskip} \epsilon_{zz}
\\ \noalign{\medskip} \epsilon_{xy}\\ \noalign{\medskip} \epsilon_{xz}\\ \noalign{\medskip} \epsilon_{zx}
\end {array} \right]; \:
\vec{\sigma} = 
\left[ \begin {array}{c} \sigma_{xx}\\ \noalign{\medskip} \sigma_{yy}\\ \noalign{\medskip} \sigma_{zz}
\\ \noalign{\medskip} \sigma_{xy}\\ \noalign{\medskip} \sigma_{xz}\\ \noalign{\medskip} \sigma_{zx}
\end {array} \right]
\end{equation}
For a general strain state, the energy density is then obtained from a simple dot-product as 
\begin{equation}
	W = \frac{1}{2} \vec{\sigma} \cdot \vec{\epsilon}.
\end{equation}
In the following, the volumetric strain energy densities for 3D and 2D plane strain will be derived.

\subsection{Pure dilatation under plane strain conditions}
In the case of pure dilatation by an amount $s$ under plane strain conditions, neither shear nor
strain along the $z$-direction is present. The corresponding strain tensor in Voigt notation is
\begin{equation}
\vec{\epsilon} = 
\left[ \begin {array}{c} s\\ \noalign{\medskip}s\\ \noalign{\medskip}0
\\ \noalign{\medskip}0\\ \noalign{\medskip}0\\ \noalign{\medskip}0
\end {array} \right] 
\end{equation}
The plane-strain energy density is therefore 
\begin{equation} \label{GCGanzenmueller::eqn:W2D}
	W_{el.}^{2D} = \frac{1}{2} \vec{\sigma} \cdot \vec{\epsilon} = \frac{9 K s^2}{4},
\end{equation}
where the fixed Poisson ratio $\nu = 1/3$, which is applicable to a 2D bond-based Peridynamic
model, has been substituted.

\subsection{Pure dilatation in 3D}
In the case of 3D pure dilatation no shear is present. Thus, 
\begin{equation}
\vec{\epsilon} = 
\left[ \begin {array}{c} s\\ \noalign{\medskip}s\\ \noalign{\medskip}s
\\ \noalign{\medskip}0\\ \noalign{\medskip}0\\ \noalign{\medskip}0
\end {array} \right], 
\end{equation}
and the volumetric energy density is
\begin{equation} \label{GCGanzenmueller::eqn:Wel3D}
	W_{el.}^{3D} = \frac{9 K s^2}{2},
\end{equation}
Note that this result is independent of $\nu$.


%
%
%

%

\begin{thebibliography}{99}
%
%
%
\bibitem{GCGanzenmueller::Silling:2000a}
S.~A.~Silling,
\emph{Reformulation of elasticity theory for discontinuities and long-range forces}
Journal of the Mechanics and Physics of Solids \textbf{48} (2000), pp.~175--209.

\bibitem{GCGanzenmueller::Silling:2005a}
S.~A.~Silling and E.~Askari,
\emph{A meshfree method based on the peridynamic model of solid mechanics},
Computers \& Structures \textbf{83} (2005), pp.~1526--1535.

\bibitem{GCGanzenmueller::Silling:2007a}
S.~A.~Silling, M.~Epton, O,~Weckner, J.~Xu, and E.~Askari,
\emph{Peridynamic States and Constitutive Modeling},
J. Elasticity \textbf{88} (2007), pp.~151--184.

\bibitem{GCGanzenmueller::Parks:2011a}
P.~Seleson and M.~L.Parks,
\emph{On the Role of the Influence Function in the Peridynamic Theory},
International Journal for Multiscale Computational Engineering, \textbf{9} (2011), pp.~689-706.

\bibitem{GCGanzenmueller::Shepard:1968a}
D.~Shepard,
\emph{A Two-dimensional Interpolation Function for Irregularly-spaced Data},
in Proceedings of the 1968 23rd ACM National Conference, New York, NY, USA, 1968, pp.~517–524.

\bibitem{GCGanzenmueller::Randles:1996a}
P.~W.~Randles and L.~D.~Libersky,
\emph{Smoothed Particle Hydrodynamics},
Computer Methods in Applied Mechanics and Engineering \textbf{139} (1996), pp.~375--408.

\bibitem{GCGanzenmueller::Kinsler:2000}
L.~E.~Kinsler~et~al.,
\emph{Fundamentals of acoustics}, 4th Ed., John Wiley and sons Inc., New York, USA, 2000.

\bibitem{GCGanzenmueller::Silling:2010a}
S.~A.~Silling, O.~Weckner, E.~Askari, and F.~Bobaru,
\emph{Crack nucleation in a peridynamic solid},
Int J Fract, \textbf{162} (2010), pp.~219--227.

\bibitem{GCGanzenmueller::Ha:2010a}
Y.~D.~Ha and F.~Bobaru,
\emph{Studies of dynamic crack propagation and crack branching with peridynamics},
Int J Fract, \textbf{162} (2010), pp.~229--244.

\bibitem{GCGanzenmueller::Ha:2011}
Y.~D.~Ha and F.~Bobaru,
\emph{Characteristics of dynamic brittle fracture captured with peridynamics},
Engineering Fracture Mechanics, \textbf{78} (2011), pp.~1156--116.

\bibitem{GCGanzenmueller::Agwai:2011a}
A.~Agwai, I.~Guven, and E.~Madenci,
\emph{Predicting crack propagation with peridynamics: a comparative study},
Int J Fract, \textbf{171} (2011), pp.~65--78.

\bibitem{GCGanzenmueller::Cook:1999}
R.~D.~Cook and W.~C.~Young,
\emph{Advanced Mechanics of Materials}, (2nd edn) Prentice-Hall, Englewood Cliffs, 1999.

\bibitem{GCGanzenmueller::VoigtNotation}
W.~Voigt,  \emph{Lehrbuch der Kristallphysik: mit Ausschlu\ss\ der Kristalloptik.}, Teubner-Verlag, 1910.

\end{thebibliography}
%

\end{document}